\documentclass[9pt, twocolumn]{article}

\usepackage{amsmath,amssymb}
\usepackage{amsxtra}
\usepackage{caption} 
 \usepackage{graphicx,epsfig}

\usepackage{float}
\usepackage{titletoc}
\usepackage{amsfonts, amscd}
\usepackage{theorem, exscale}
\usepackage{epic, eepic}
\usepackage{epsfig}
\usepackage{hyperref} %
\usepackage[usenames]{color}

\usepackage{makeidx}

\usepackage[normalem]{ulem}

\makeindex

\newcounter{Figure}
\theoremstyle{plain}
\theoremheaderfont{\scshape}

\newtheorem{Def}{\bf Definition}
\newtheorem{The}{\bf Theorem}

\theorembodyfont{\upshape}

\theoremheaderfont{\scshape\small} \theorembodyfont{\scshape\small}

\newcommand{\real}{ {\mathbb R} }

\newcommand{\be}{\begin{equation}}
\newcommand{\ee}{\end{equation}}
\newcommand{\bea}{\begin{eqnarray}}
\newcommand{\eea}{\end{eqnarray}}
\newcommand{\beas}{\begin{eqnarray*}}
\newcommand{\eeas}{\end{eqnarray*}}

\begin{document}

\begin{center}
{\Large \bf Gravitational Waves \\ 
\vspace{3pt} and Their Mathematics } \\ 
\end{center}
\vspace{0.1cm}

\begin{center}
{\large \bf Lydia Bieri, David Garfinkle, Nicol\'as Yunes} \\ 
\end{center}

\vspace{0.1cm}

{\small \scshape A version of this article was published in the AMS Notices, Vol. 64, Issue 07, 2017.} 
\footnote{A version of this article (with more graphics and photos) was published in the AMS Notices, Vol. 64, Issue 07, 2017, (August issue 2017). 
It is available for free to download from the journal's website under 
\url{http://www.ams.org/publications/journals/notices/201707/rnoti-p693.pdf} \\ 
The full August 2017 issue is available under 
\url{http://www.ams.org/journals/notices/201707/index.html}}

\vspace{.5cm}

\section{Introduction} 
\label{intro}
%

In 2015 gravitational waves were detected for the first time by the LIGO-Virgo collaboration \cite{ligo1}. This triumph happened 
100 years after Albert Einstein's formulation of the Theory of General Relativity \cite{ae1915*1, ae1915*2, ae1916*1} and 99 years after his prediction of gravitational waves \cite{ae1916*2}.  
 This article focuses on the mathematics of Einstein's gravitational waves, from the properties of the Einstein vacuum equations and the initial value problem (Cauchy problem), to the various approximations used to obtain quantitative predictions from these equations, and eventually an experimental detection.

\begin{figure}[htb]
\begin{center}
\includegraphics[width=0.4\textwidth]{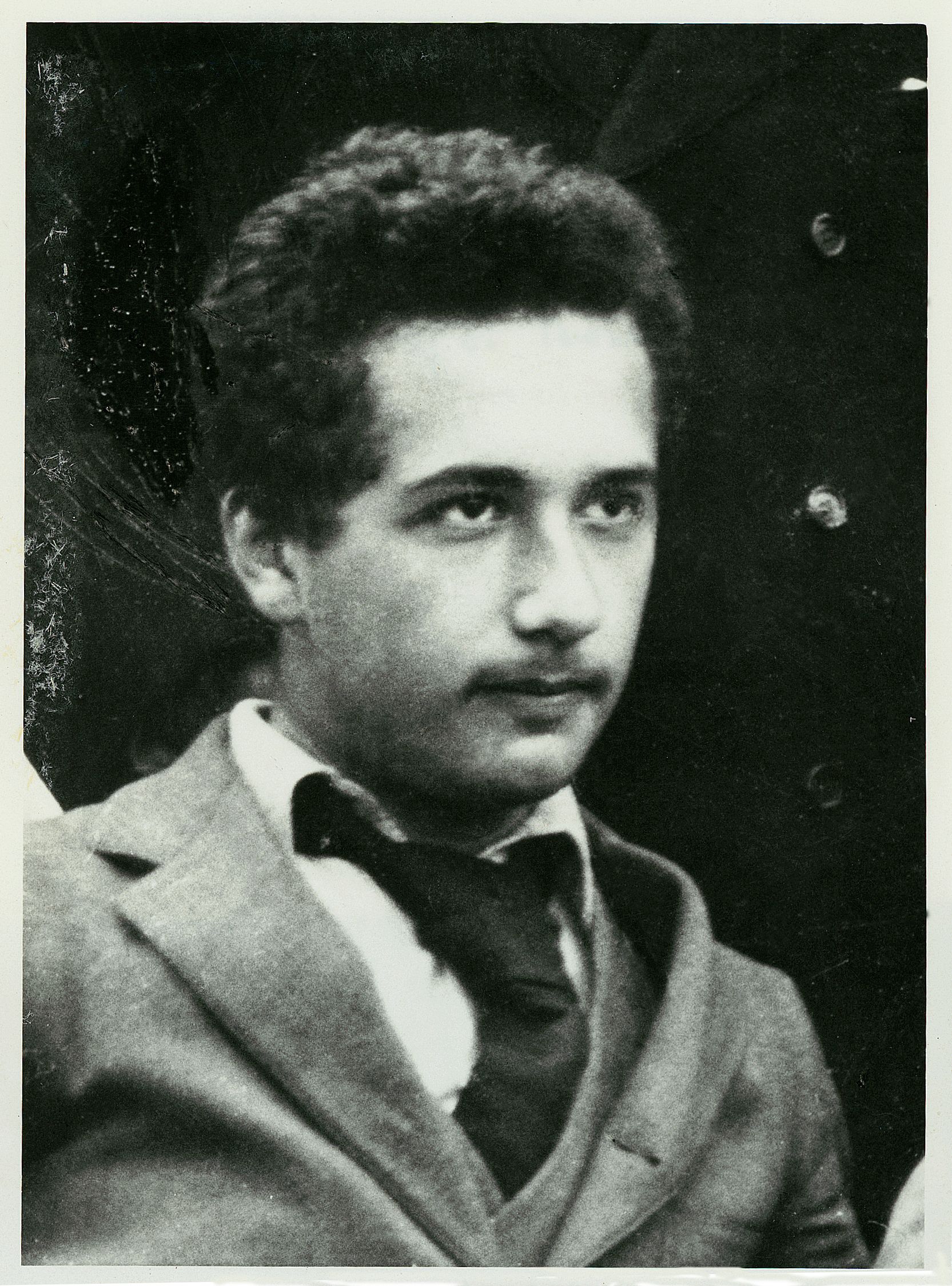}
\caption{Albert Einstein predicted gravitational waves in 1916.}
\label{Einstein}
\end{center}
\end{figure}

General Relativity is studied as a branch of astronomy, physics, and mathematics. At its core are the Einstein equations, which link the physical content of our universe to geometry. By solving these equations, we construct the spacetime itself, a continuum that relates space, time, geometry, and matter (including energy). The dynamics of the gravitational field are studied in the Cauchy problem for the Einstein equations, relying on the theory of nonlinear partial differential equations (pde) and geometric analysis. The connections between astronomy, physics, and mathematics are richly illustrated by the story of gravitational radiation. 

In General Relativity, the universe is described as a spacetime manifold with a curved metric whose curvature encodes the properties of the gravitational field.  While sometimes one wants to use General Relativity to describe the whole universe, often we just want to know how a single object or small collection of objects behaves.  To address that kind of problem, we use the idealization of the isolated system: a spacetime consisting of just the objects we want to study and nothing else.  We might consider the solar system as an isolated object, or a pair of black holes spiraling into one another until they collide.  We ask how those objects look to a distant, far away observer in a region where presumably the curvature of spacetime is very small.  Gravitational waves are vibrations in spacetime that propagate at the speed of light away from their source.  They may be produced, for example, when black holes merge.  This is what was first detected by Advanced LIGO (aLIGO) and this is the focus of this article.

First we describe the basic differential geometry used to define the universe as a geometric object.  Next we describe the mathematical properties of the Einstein vacuum equations, including a discussion of the Cauchy problem and
gravitational radiation. Then we turn to the various approximation schemes used to obtain quantitative predictions from these equations.
We conclude with the experimental detection of gravitational waves and the astrophysical implications of this detection. 
This detection is not only a spectacular confirmation of Einstein's theory, but also the beginning of the era of gravitational wave astronomy, the use of gravitational waves to investigate aspects of our universe that have been inaccessible to telescopes.  \\

\section{The Universe as a Geometric Object}
\label{geometry}

A spacetime manifold is defined to be a $4$-dimensional, oriented, differentiable manifold $M$ with a 
Lorentzian metric tensor, $g$, which is 
a non-degenerate quadratic form of index one,
$$
g =\sum_{\mu, \nu=0}^3 g_{\mu\nu} dx^\mu\otimes dx^\nu,
$$
defined in $T_qM$ for every $q$ in $M$ varying smoothly in $q$. 
The trivial example, the Minkowski spacetime as defined in Einstein's Special Relativity,
is ${\mathbb{R}}^4$ endowed with the flat Minkowski metric: 
\be\label{Mink-here}
g=\eta= - c^2 dt^2 + dx^2+dy^2+dz^2.
\ee
Taking $x_0=t$, $x_1=x$, $x_2=y$ and $x_3=z$, we have
 $\eta_{0 0}=-c^2$, $\eta_{ii}=1$ for $i=1,2,3$ and $\eta_{\mu\nu}=0$ for $\mu\neq\nu$. In mathematical General Relativity we often normalize the 
 speed of light, $c=1$.

The family of Schwarzschild metrics are solutions of the Einstein vacuum equations that describe spacetimes containing a black hole, where the parameter values are $M > 0$.  Taking
 $r_s=2GM/c^2$, it has the metric:
\be \label{Sch-here}
g=- c^2 \frac{\left(1-\tfrac{r_s}{4\rho}\right)^2}{\left(1+\tfrac{r_s}{4\rho}\right)^2}\,dt^2
+ \left(1+\tfrac{r_s}{4\rho}\right)^4 \, h  , 
\ee
where $\rho^2=x^2+y^2+z^2$, 
$h =d{x^2}+d{y^2}+d{z^2}$, and $G$ denotes the Newtonian gravitational constant. This space is asymptotically flat as $\rho\to \infty$.  

The~Friedmann-Lema\^itre-Robertson-Walker spacetimes describe homogeneous and isotropic
universes through the metric
\be \label{FLRW}
g=-c^2 dt^2 + a^2(t) g_{\chi} , 
\ee
where $g_{\chi}$ is a Riemannian metric with constant sectional
curvature, $\chi$, (e.g. a sphere when $\chi=1$) and $a(t)$ describes the expansion of the universe.   The function, $a(t)$, is found by solving the Einstein equations as sourced by fluid matter. 
  
In an arbitrary Lorentzian manifold, $M$, a vector $X \in T_xM$ is called 
{\em null} or lightlike if 
$$g_x (X, X) = 0.$$ 
At every point there is a cone of null vectors called the {\em null cone}. 
A vector $X \in T_xM$ is called
{\em timelike} if 
$$g_x (X, X)  <  0,$$ 
and {\em spacelike} if 
$$g_x (X, X) > 0.$$  
In General Relativity nothing travels faster than the speed of light, 
so the velocities of massless particles are null vectors whereas those for massive objects are timelike.   A causal curve is a differentiable curve for which the tangent vector at each point is either timelike or null.

A hypersurface is called {\em spacelike} if its normal vector is timelike, so that the
metric tensor restricted to the hypersurface is positive definite.   A {\em Cauchy hypersurface} is a spacelike hypersurface where each causal curve through any point $x \in M$ intersects $\mathcal{H}$ exactly at one point.  
A spacetime $(M,g)$ is said to be {\em globally hyperbolic} if it has a Cauchy hypersurface. 
In a globally hyperbolic spacetime, there is a time function $t$ whose gradient is everywhere timelike or null and whose level surfaces are Cauchy surfaces.   A globally hyperbolic spacetime is causal in the sense that no object may travel to its own past.

As in Riemannian geometry, curves with $0$ acceleration are called geodesics.   Light travels along null geodesics.  Geodesics which enter the event horizon of 
a black hole never leave.  Objects in free fall travel along timelike geodesics.   They
also can never leave once they have entered a black hole.  When two black holes
fall into one another, they merge and form a single larger black hole.

In curved spacetime, geodesics bend together or apart and the relative
acceleration between geodesics is described by the Jacobi equation, also known as the geodesic deviation equation. In particular, the 
relative acceleration of nearby geodesics is given by the Riemann curvature tensor times the distance between them.  The Ricci curvature tensor, $R_{\mu\nu}$, measures the average way in which geodesics curve together or apart. The scalar curvature, $R$, is the trace of the Ricci curvature.  

Einstein's field equations are:
\be \label{einsta1}
R_{\mu \nu} - {\textstyle {\frac{1}{2}}} R g_{\mu \nu} = \frac{8 \pi G}{c^4} T_{\mu \nu} \ , 
\ee
where $T_{\mu \nu}$ denotes the energy-momentum tensor, which encodes the energy density of matter. 
Note that for cosmological considerations, one can add $\Lambda g_{\mu\nu}$ on the left hand side, where $\Lambda$ is the cosmological constant. However, nowadays, this term is commonly absorbed into $T_{\mu \nu}$ on the right hand side. 
Here we will consider the non-cosmological setting. One then solves the Einstein equations for the metric tensor $g_{\mu \nu}$.   If there are no other fields, then $T_{\mu \nu}=0$ and (\ref{einsta1}) reduce to the Einstein vacuum equations: 
\be \label{einsta3}
R_{\mu \nu} = 0 \ . 
\ee

Note that the Einstein Equation is a set of second order quasilinear partial differential equations for the metric tensor.  
In fact, when choosing the right coordinate chart (wave coordinates), taking $c=1$, and writing out the formula for the
curvature tensor, $R_{\mu\nu}$, in those coordinates, the equation becomes:
\be \label{redg1}
\Box_g g_{\alpha \beta} = N_{\alpha \beta} 
\ee
where $\Box_g$ is the wave operator and $N_{\alpha \beta} = N_{\alpha \beta} (g, \partial g)$ denote nonlinear terms with quadratics in $\partial g$. 

Quite a few exact solutions to the Einstein vacuum equations are known. Among the most popular are the trivial solution  
(Minkowski spacetime) as in (\ref{Mink-here}), the Schwarzschild solution, which describes a static black hole, as in (\ref{Sch-here}), and the Kerr solution, which describes a black hole with spin angular momentum.  Note that the exterior gravitational field of any spherically symmetric object  
takes the form of (\ref{Sch-here}) for $r>r_0$ where $r_0>r_s$ is the radius of the object, so this model can
be used to study the spacetime around an isolated star or planet. 
However, in order to understand the dynamics of the gravitational field and radiation, we have to investigate large classes of spacetimes. This can only be done by solving the initial value problem (Cauchy problem) for the Einstein equations, which will be discussed in the next section.

If there are matter fields, so that $T_{\mu \nu} \neq 0$, then these fields satisfy their own evolution equations, which have to be solved along with the Einstein field equations (\ref{einsta1}) as a coupled system.   The scale factor, $a(t)$, of the Friedmann-Lema\^itre-Robertson-Walker cosmological spacetimes in (\ref{FLRW}) can then be found by solving a second order, ordinary differential equation derived from (\ref{einsta1}).   If a solution has a time where the scale factor vanishes, then the solution is said to describe a cosmos whose early phase is a ``big bang." 

\begin{figure}[htb]
\begin{center}
\includegraphics[width=0.4\textwidth]{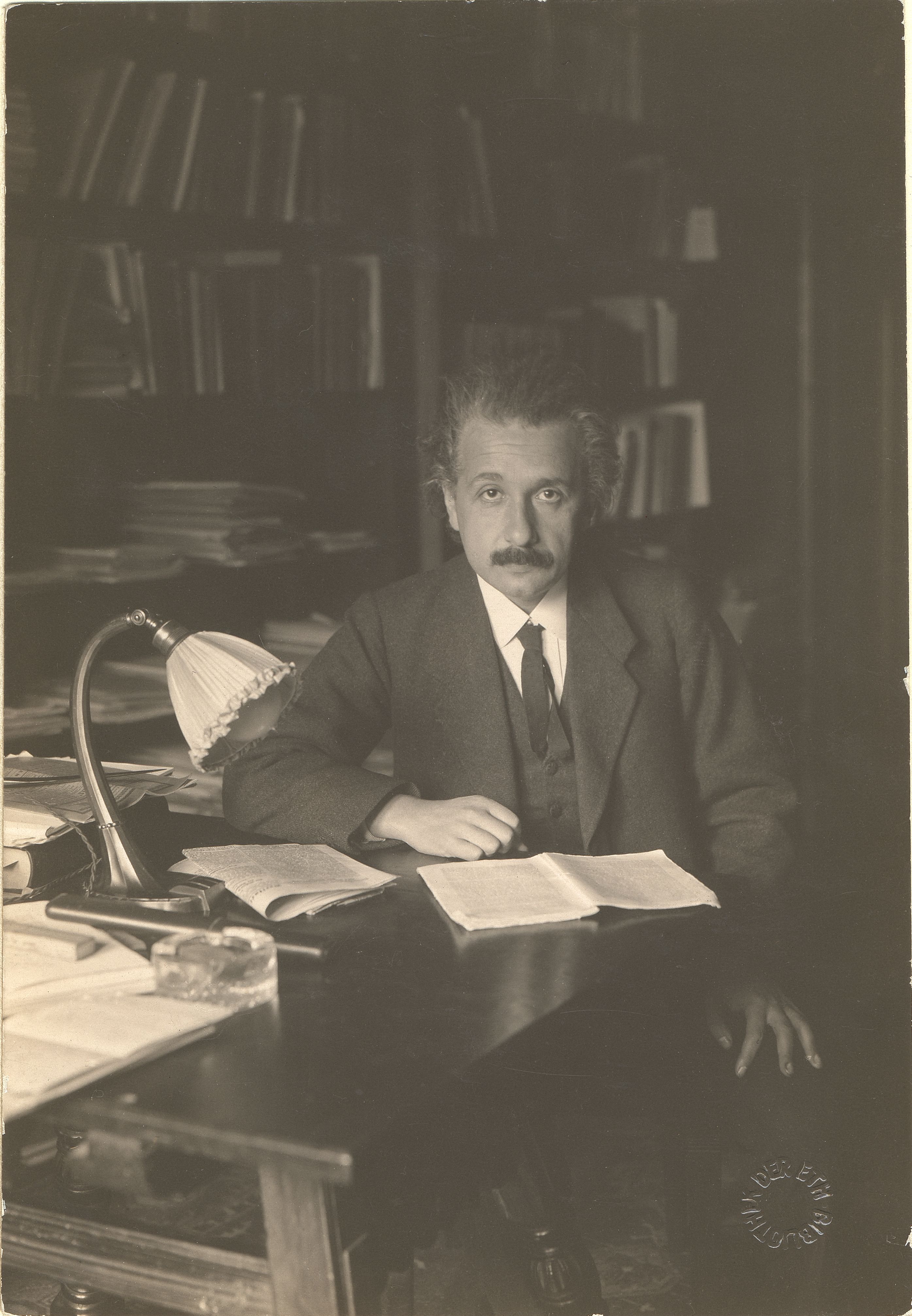}
\caption{Albert Einstein}
\label{Einstein2}
\end{center}
\end{figure}

\section{The Einstein Equations}

\subsection{Beginnings of Cauchy Problem} 
\label{cauchy}

In order to study gravitational waves, stability problems, and general questions about the dynamics of the gravitational field, we have to formulate and solve the {\em Cauchy problem}.   That is we are given initial data: a prescribed Riemannian manifold $\mathcal{H}$ with a complete Riemannian metric $\bar{g}_{ij}$ and a symmetric $2$-tensor $\mathcal{K}_{ij}$ satisfying certain consistency
 conditions called the {\em Einstein constraint equations}.   We then solve for a spacetime $(M,g)$ which satisfies the Einstein equations evolving forward from this initial data set.  That is, the given Riemannian manifold $\mathcal{H}$ 
 is a spacelike hypersurface in this spacetime solution $M$, where $\bar{g}$ is the restriction of $g$.  Furthermore, the symmetric two tensor $\mathcal{K}_{ij}$ is the prescribed second fundamental form.   
 
All the different methods used to describe gravitational radiation have to be thought of as embedded into the aim of solving the Cauchy problem. We solve the Cauchy problem by methods of analysis and geometry. However, for situations where the geometric-analytic techniques are not (yet) at hand, one uses approximation methods and numerical algorithms. The goal of the latter methods is to produce approximations to solutions of the Cauchy problem for the Einstein equations.  

In order to derive the gravitational waves from binary black hole mergers, binary neutron star mergers, or core-collapse supernovae, we describe these systems by asymptotically flat spacetimes. These are 
solutions of the Einstein equations that at infinity tend to Minkowski space
with a metric as in (\ref{Mink-here}).  Schwarzschild space is a simple example of such an isolated system containing only a single stationary black hole (\ref{Sch-here}).  There is a huge literature about specific fall-off rates which we will not
describe here.  The null asymptotics of these spacetimes contain information on gravitational radiation (gravitational waves) out to infinity.

Recall that the Einstein vacuum equations (\ref{einsta3}) are a system of ten quasilinear, partial differential equations that can be put into hyperbolic form. However, with the Bianchi identity imposing four constraints, the Einstein vacuum system (\ref{einsta3}) constitutes only six independent equations for the ten unknowns of the metric $g_{\mu \nu}$. This corresponds to the general covariance of the Einstein equations. In fact, uniqueness of solutions to these equations holds up to equivalence under diffeomorphisms. We have just found a core feature of General Relativity. This mathematical fact also means that physical laws do not depend on the coordinates used to describe a particular process.

The Einstein equations split into a set of evolution equations and a set of constraint equations. 
As above, $t$ denotes the time coordinate whereas indices $i, j = 1, \cdots , 3$ refer to spatial coordinates.  Taking $c=1$, the evolution equations read: 
\be
\frac{\partial \bar{g}_{ij}}{\partial t} \  =  \ - 2 \Phi \mathcal{K}_{ij} + \mathcal{L}_X \bar{g}_{ij}  \label{evol1}
\ee
\bea
\frac{\partial \mathcal{K}_{ij}}{\partial t} \  =  \  - \nabla_i  \nabla_j \Phi  + \mathcal{L}_X \mathcal{K}_{ij}\nonumber \\  
\  + \ (\bar{R}_{ij} \ + \ \mathcal{K}_{ij} \ tr \mathcal{K} \  - \ 2 \mathcal{K}_{im} \mathcal{K}^m_j) \Phi    
\label{evol2}
\eea
Here $\mathcal{K}_{ij}$ is the extrinsic curvature of the $t=const.$ surface $\mathcal{H}$ as above.  The lapse $\Phi$ and shift $X$ are essentially the $g_{tt}$ and $g_{ti}$ components of the metric, and are given by 
$T={\Phi}n +X$ where $T$ is the evolution vector field $\partial /\partial t$ and $n$ is the unit normal to the constant time hypersurface.  $\nabla _i$ is the spatial covariant derivative and $\mathcal{L}$ is the Lie derivative. 
However, the initial data $({\bar{g}_{ij}},{{\mathcal K}_{ij}})$ cannot be chosen freely: the remaining four Einstein vacuum equations become the following constraint equations: 
\bea
\nabla^i  \mathcal{K}_{ij} \ - \ \nabla_j \ tr \mathcal{K} \ &  = & \ 0   \label{con1} \\
\bar{R} \ + \ (tr \mathcal{K})^2 \ - \ | \mathcal{K} |^2 \ & = & \ 0 \ . \label{con2}
\eea

An {\itshape initial data set} is a $3$-dimensional manifold $\mathcal{H}$ with a complete Riemannian metric $\bar{g}_{ij}$ and a symmetric $2$-tensor $\mathcal{K}_{ij}$ satisfying the constraint equations ((\ref{con1}), (\ref{con2})). 
We will evolve an asymptotically flat initial data set $(\mathcal{H}, \bar{g}_{ij}, \mathcal{K}_{ij})$, that outside a sufficiently large compact set $\mathcal{D}$, $\mathcal{H} \backslash \mathcal{D}$ is diffeomorphic to the complement of a closed ball in $\real^3$ and admits a system of coordinates where $\bar{g}_{ij} \to \delta_{ij}$ and $\mathcal{K}_{ij} \to 0$ sufficiently fast. 

It took a long time before the Cauchy problem for the Einstein equations was formulated correctly and understood. Geometry and pde theory were not as developed as they are today, and the pioneers of General Relativity had to struggle with problems that have elegant solutions nowadays. 
The beauty and challenges of General Relativity attracted many mathematicians, as for instance D. Hilbert or H. Weyl, to work on General Relativity's fundamental questions. 
Weyl in 1923 talked about a ``causally connected" world, which hints at issues that the domain of dependence theorem much later would solve. 
G. Darmois in the 1920s studied the analytic case, which is not physical but a step in the right direction. He recognized that the analyticity hypothesis is physically unsatisfactory, because it hides the propagation properties of the gravitational field. Without going into details, important work followed by K. Stellmacher, K. Friedrichs, T. de Donder, and C. Lanczos. The latter two introduced wave coordinates, which Darmois later used. In 1939, A. Lichnerowicz extended Darmois' work. He also suggested the extension of the $3+1$ decomposition with non-zero shift to his student Yvonne Choquet-Bruhat, which she carried out.

Choquet-Bruhat, encouraged by Jean Leray in 1947, searched for a solution to the non-analytic Cauchy problem of the Einstein equations, which turned into her famous result of 1952. There are many more players in this game that should be mentioned, but there is not enough space to do justice to their work. These works also built on progress in analysis and pde theory by H. Lewy, J.  Hadamard, J. Schauder, and S. Sobolev among many others.   Details on the history of the proof can be found in Choquet-Bruhat's survey article published in \cite{ycbsurveyjdg1} {\itshape Surveys in Differential Geometry 2015: One hundred years of general relativity}, and more historical background (including a discussion between Choquet-Bruhat and Einstein) is given in Choquet-Bruhat's forthcoming autobiography. 

In 1952 Choquet-Bruhat \cite{bru} proved a local existence and uniqueness theorem for the Einstein equations, and in 1969 Choquet-Bruhat and R.  Geroch \cite{bruger} proved the global existence of a unique maximal future development for every given initial data set. 

\begin{The} \label{ycb1952} (Choquet-Bruhat, 1952)
Let $(\mathcal{H}, \bar{g}, \mathcal{K})$ be an initial data set satisfying the vacuum constraint equations. Then there exists a spacetime $(M, g)$ satisfying the Einstein vacuum equations with $\mathcal{H} \hookrightarrow M$ being a spacelike surface with induced metric $\bar{g}$ and second fundamental form $\mathcal{K}$. 
\end{The}

This was proven by finding a useful coordinate system, called wave coordinates, in which Einstein's vacuum equations appear clearly as a hyperbolic system of
partial differential equations.  The pioneering result by Choquet-Bruhat was improved by Dionne (1962), Fisher-Marsden (1970), and Hughes-Kato-Marsden (1977) using the energy method.   

\subsection{Global Cauchy Problem}
\label{globalCauchy}

Choquet-Bruhat's local theorem \ref{ycb1952} of 1952 was a breakthrough and has since been fundamental for further investigations of the Cauchy problem. 
Once we have local solutions of the Einstein equations, do they exist for all time, or do they form singularities? And of what type would the latter be? In 1969
Choquet-Bruhat and Geroch proved there exists a unique, globally hyperbolic, maximal spacetime $(M, g)$ satisfying the Einstein vacuum equations with $\mathcal{H} \hookrightarrow M$ being a Cauchy surface with induced metric $\bar{g}$ and second fundamental form $\mathcal{K}$.   This unique solution is called the
 {\em maximal future development} of the initial data set. 

However, there is no information about the behavior of the solution. 
Will singularities occur or will it be complete? 
One would expect that sufficiently small initial data evolves forever without producing any singularities, whereas sufficiently large data evolves to form spacetime singularities such as black holes. From a mathematical point of view the question is whether theorems can be proven that establish this behavior. A breakthrough occurred in 2008 with Christodoulou's proof \cite{chrdmay2008}, building on an earlier result due to Penrose \cite{pen2}, that black hole singularities form in the Cauchy development of initial data, which do not contain any singularities, provided that the incoming energy per unit solid angle in each direction in a suitably small time interval is sufficiently large. This means that a black hole forms through the focussing of gravitational waves. This result has since been generalized by various authors, and the main methods have been applied to other nonlinear pdes.

The next burning question to ask is whether there is any asymptotically flat (and non-trivial) initial data with complete maximal development.  This can be thought of as a question about the global stability of Minkowski space. In their celebrated work \cite{sta} of 1993 D. Christodoulou and S. Klainerman proved the following result, which here we state in a very general way. The details are intricate and the smallness assumptions are stated for weighted Sobolev norms of the geometric quantities. 

\begin{The} \label{chrddklais1993} (Christodoulou and Klainerman, 1993, \cite{sta}) 
Given strongly asymptotically flat initial data for the Einstein vacuum equations (\ref{einsta3}), which is sufficiently small, there exists a unique, causally geodesically complete and globally hyperbolic solution $(M, g)$, which itself is globally asymptotically flat. 
\end{The}

The proof relies on geometric analysis and is independent of coordinates. First, energies are identified with the help of the Bel-Robinson tensor, which basically is a quadratic of the Weyl curvature. Then, the 
curvature components are estimated in a comparison argument using the energies. Finally, in a large bootstrap argument with assumptions on the curvature, the remaining geometric quantities are proven to be controlled. The proof comprises various new ideas and features that became important not only for further studies of relativistic problems but also in other nonlinear hyperbolic pdes.

The Christodoulou-Klainerman result of theorem \ref{chrddklais1993} was generalized in 2000 by Nina Zipser \cite{zip}, \cite{zip2} for the Einstein-Maxwell equations and in 2007 by Lydia Bieri \cite{lydia1}, \cite{lydia2} for the Einstein vacuum equations assuming less on the decay at infinity and less regularity. Thus, the latter result establishes the borderline case for decay of initial data in the Einstein vacuum case. Both works use geometric analysis in a way that is independent of any coordinates.  

Next, let us go back to the pioneering results by Choquet-Bruhat and Geroch, and say a few words about further extensions of these works.
A standard result ensures that for an Einstein vacuum initial data set $(\mathcal{H}_0, \bar{g}, \mathcal{K})$ with $\mathcal{H}_0$ allowing to be covered by a locally finite system of coordinate charts with transformations being $C^1$-diffeomorphisms, and 
\be
g_{mn} |_{\mathcal{H}_0} \in H^{k}_{loc} \ \ , \ \ \ \partial_0 g_{mn} |_{\mathcal{H}_0} \in H^{k-1}_{loc} \ \ , \ \ \ k > \frac{5}{2} \ , 
\ee
there exists a unique globally hyperbolic solution with $\mathcal{H}_0$ being a Cauchy hypersurface. 
Several improvements followed, including those by Bahouri-Chemin \cite{bach}, Tataru \cite{tat1}, Smith-Tataru \cite{stat1}, and then by Klainerman-Rodnianski \cite{rkr1}, \cite{rkr2}. 
The latter proved that for the same problem but with $k > 2$ there exists a time interval $[0,T]$ and a unique solution $g$ such that 
$g_{mn} \in C^0([0,T], H^k)$ where $T$ depends only on $|| g_{mn} |_{\mathcal{H}_0} ||_{H^k} + || \partial_0 g_{mn} |_{\mathcal{H}_0} ||_{H^{k-1}}$. Recently the $L^2$ curvature conjecture was proven by Klainerman-Rodnianski-Szeftel \cite{kleinrod2}, \cite{kleinrod3}, \cite{sze1}, \cite{sze2}, \cite{sze3}, \cite{sze4}: under certain assumptions they relax the regularity condition such that the time of existence of the solution depends only on the $L^2$-norms of the Riemann curvature tensor and on the gradient of the second fundamental form. 

For our purposes, we want to know what the global existence theorem says about the properties of radiation, i.e.~the behavior of curvature at large distances.  In particular, because we expect gravitational radiation to propagate at the speed of light, we would like to study the behavior at large distances along outgoing light rays.  This sort of question was addressed long before Christodoulou and Klainerman. However, these works assume a lot on the spacetimes considered. As a consequence, components of the Riemann curvature tensor show a specific hierarchy of decay in $r$. The spacetimes of the Christodoulou-Klainerman theorem do not fully satisfy these properties, showing only some of the fall-off but not all. In fact, Christodoulou showed that physical spacetimes cannot fulfill the stronger decay. The results by Christodoulou-Klainerman provide a precise description of null infinity for physically interesting situations.

\subsection{Gravitational Radiation}
\label{radiation}

In this section, we consider radiative spacetimes with asymptotic structures as derived by Christodoulou-Klainerman \cite{sta}.   The asymptotic behavior of gravitational waves near infinity~\cite{BBM,sachs,pen,gerochaf1,gerochaf2} approximates how gravitational radiation emanating from a distant black hole merger would appear when observed by aLIGO.  Asymptotically the gravitational waves appear to be planar, stretching and shrinking directions perpendicular to the wave's travel direction. 

As an example, let us consider the merger of two black holes. Long before the merger, the total energy of the two-black-hole spacetime, the so-called ADM energy or ``mass," named for its creators Arnowitt-Deser-Misner~\cite{adm}, is essentially the sum of the masses of the individual black holes. During the merger, energy and momentum are radiated away in the form of gravitational waves. After the merger, once the waves have propagated away from the system, the energy left in the system, the so-called Bondi mass, decreases and this can be calculated through the formalism introduced by Bondi, Sachs, and Trautman.

Gravitational radiation travels along null hypersurfaces in the spacetime. As the source is very far away from us, we can think of these waves as reaching us (the experiment) at null infinity, which is defined as follows.
\begin{Def}
{\bf Future null infinity} $\mathcal{I}^+$ is defined to be the endpoints of all future-directed null geodesics along which $r \to \infty$. It has the topology of 
$\real \times \mathbb{S}^2$ with the function $u$ taking values in $\real$. 
\end{Def}
A null hypersurface $\mathcal{C}_u$ intersects $\mathcal{I}^+$ at infinity in a $2$-sphere. To each $\mathcal{C}_u$ at null infinity is assigned a Trautman-Bondi mass $M(u)$, as introduced by Bondi, Trautman, and Sachs in the middle of the last century. This 
quantity measures the amount of mass that remains in an isolated gravitational system at a given retarded time, i.e.~the Trautman-Bondi mass measures the remaining mass after radiation through $\mathcal{I}^+$ up to $u$. 
The {Bondi mass-loss formula} reads for $u_1 \leq u_2$ 
\be \label{Bondimassloss}
M\left( u_2 \right) 
 =
 M\left( u_1 \right) - 
C 
 \int_{u_1}^{u_2} 
\int_{S^{2}} \left| \Xi \right| ^{2} d\mu _{\overset{\circ }{\gamma }} du
\ee
with $ \left| \Xi \right| ^{2}$ being the norm of the shear tensor at $\mathcal{I}^+$ and $d\mu _{\overset{\circ }{\gamma }}$ the canonical measure on $S^2$. 
If other fields are present, like electromagnetic fields, then the formula contains a corresponding term for that field. 
In the situations considered here, it has been proven that 
$\lim_{u \to - \infty} M(u) = M_{ADM}$. 

The effects of gravitational waves on neighboring geodesics is encoded in the Jacobi equation. This very fact is at the heart of the detection by aLIGO and is discussed in Sec.~\ref{experiment}. From this, we derive a formula for the displacement of test masses, while the wave packet is traveling through the apparatus. This is what was measured by the aLIGO detectors.

Now, there is more to the story. From the analysis of the spacetime at 
$\mathcal{I}^+$ one can prove that the test masses will go to rest after the gravitational wave has passed, meaning that the geodesics will not be deviated anymore. However, will the test masses be at the ``same" position as before the wave train passed or will they be dislocated? In mathematical language, will the spacetime geometry have changed permanently?  If so, then this is called the {\em memory effect} of gravitational waves. This effect was first computed in 1974 by Ya.B. Zel'dovich and A.G. Polnarev in the linearized theory \cite{zelpol}, where it was found to be very small and considered not detectable at that time.

In 1991 D. Christodoulou, studying the full nonlinear problem \cite{chrmemory}, showed that this effect is larger than expected and could in principle be measured. Bieri and Garfinkle showed \cite{lbdg3} that the formerly called ``linear" (now ordinary) and ``nonlinear" (now null) memories are two different effects, the former sourced by the difference of a specific component of the Weyl tensor, and the latter due to fields that do reach null infinity $\mathcal{I}^+$.
In the case of the Einstein vacuum equations, this is the shear appearing in (\ref{Bondimassloss}). In particular, the permanent displacement (memory) is related to 
\be \label{energy1}
\mathcal{F} = C \int_{- \infty}^{+ \infty}  \left| \Xi (u ) \right| ^{2} du 
\ee
where $\mathcal{F}/4 \pi$ denotes the total energy radiated in a given direction per unit solid angle. 
A very recent paper by P.D. Lasky, E. Thrane, Y. Levin, J. Blackman, and Y. Chen \cite{laskyetal} suggests a method for detecting gravitational wave memory with aLIGO.

\section{Approximation Methods}
\label{wavesdetail}

To compare gravitational wave experimental data to the predictions of the theory, one needs a calculation of the predictions of the theory.  It is not enough to know that solutions of the Einstein field equations exist; rather, one needs quantitative solutions of those equations to at least the accuracy needed to compare to experiments.
In addition, sometimes the gravitational wave signal is so weak that to keep it from being overwhelmed by noise one must use the technique of matched filtering~\cite{Jaranowski:2007pe} in which one looks for matches between the signal and a set of templates of possible expected waveforms.  These quantitative solutions are provided by a set of overlapping approximation techniques, and by numerical simulations.  We will discuss the approximation techniques in this section and the numerical methods in section \ref{mathnum}.

\subsubsection{Linearized Theory \\ and Gravitational Waves}

Since gravitational waves become weaker as they propagate away from their sources, one might hope to neglect the nonlinearities of the Einstein field equations
and focus instead on the linearized equations, which are easier to work with.  One
may hope that these equations would provide an approximate description
of the gravitational radiation for much of its propagation and for its interaction with the detector.  
In linearized gravity, one then writes the spacetime metric as
\be \label{lin1}
g_{\mu \nu} = \eta_{\mu \nu} + h_{\mu \nu} 
\ee
where $\eta_{\mu \nu}$ is the Minkowski metric as in (\ref{Mink-here})
and $h_{\mu \nu}$ is assumed to be small.  One then keeps terms in the Einstein field equations only to linear order in $h_{\mu \nu}$.  The coordinate invariance of General Relativity gives rise to what is called gauge invariance in linearized gravity.  In particular, consider any quantity $F$ written as $F = {\bar F} + \delta F$ where $\bar F$ is the value of the quantity in the background and $\delta F$ is the first order perturbation of that quantity.  Then for an infinitesimal diffeomorphism along the vector field $\xi$, the quantity $\delta F$ changes by 
$$
\delta F \to \delta F + {{\cal L}_\xi}{\bar F},
$$ 
where recall that $\cal L$ stands for the Lie derivative.  Recall also that harmonic coordinates made the Einstein vacuum equations look like the wave equation in (\ref{redg1}).  
We would like to do something similar in linearized gravity.  To this end we choose $\xi$ to impose the Lorenz gauge condition (not Lorentz!) 
$$\partial_{\mu} \bar{h}^{\mu \nu} = 0,$$ where $$\bar{h}_{\mu \nu} = h_{\mu \nu} - (1/2) \eta_{\mu \nu} h.$$  The linearized Einstein field equations then become
\be
\Box {{\bar h}_{\mu \nu}} = - 16 \pi G {T_{\mu \nu}} , 
\ee
where $\Box$ is the wave operator in Minkowski spacetime.

In a vacuum one can use the remaining freedom to choose $\xi$ to impose the conditions that $h_{\mu \nu}$ has only spatial components and is trace-free, while remaining in Lorenz gauge.  
This refinement of the Lorenz gauge is called the TT gauge, since it guarantees that the only two propagating degrees of freedom of the metric perturbation are transverse $\partial^{i} h_{ij}=0$ and (spatially) traceless $\eta^{ij} h_{ij} = 0$.  The metric in TT gauge has a direct physical interpretation given by the following 
formula for the linearized Riemannian curvature tensor 
\be \label{approxR1}
R_{itjt} = - \frac{1}{2} \ddot{h}^{TT}_{ij}\,,
\ee
which sources the geodesic deviation equation, and thus encapsulates how matter behaves in the presence of gravitational waves. Combining Eq.~(\ref{approxR1}) and the Jacobi equation, one can compute the change in distance between 
two test masses in free fall~\cite{Pitkin:2011yk}: 
\[
\triangle d^i (t) = \frac{1}{2} h^{TT}_{ij} (t) d_0^j
\]
where $d_0^{j}$ is the initial distance between the test masses. 

The TT nature of gravitational wave perturbations allows us to immediately infer that they only have two polarizations. Consider a wave traveling along the $z$-direction, such that $h^{TT}_{ij} (t-z)$ is a solution of $\Box h^{TT}_{ij} = 0$. The Lorenz condition, the assumption that the metric perturbation vanishes for large $r$, the trace-free condition, and symmetries imply that there are only two independent propagating degrees of freedom: 
$$
h_+ (t-z) = h^{TT}_{xx} = - h^{TT}_{yy}
$$
and 
$$
h_\times (t-z)= h^{TT}_{xy} = h^{TT}_{yx}.  
$$
The $h_+$ gravitational wave stretches the $x$ direction in space while it squeezes the $y$ direction, and vice-versa.
The interferometer used to detect gravitational waves has two long perpendicular arms that measure this distortion.   
Therefore, one must approximate these displacements
in order to predict what the interferometer will see under various scenarios.

\subsection{The post-Newtonian Approximation}

The post-Newtonian (PN) approximation for gravitational waves~\cite{Blanchet:2006zz} extends the linearized study presented above to higher orders in the metric perturbation, while also assuming that the bodies generating the gravitational field move slowly compared to the speed of light~\cite{PoissonWill}. The PN approach was developed by Einstein, Infeld, Hoffman, Damour, Deruelle, Blanchet, Will, Schaefer, and many others (see~\cite{Blanchet:2006zz,PoissonWill} and references therein). In the harmonic gauge $\partial_{\alpha} (\sqrt{-g} g^{\alpha \beta}) = 0$ commonly employed in PN theory, the expanded equations take the form
\be \label{rel}
\Box h^{\alpha \beta} = - \frac{16 \pi G}{c^4} \tau^{\alpha \beta}\,,
\ee
where $\Box$ is the wave operator and
$$
\tau^{\alpha \beta} = -(g) T^{\alpha \beta} + (16 \pi)^{-1} N^{\alpha \beta},
$$
with $N^{\alpha \beta}$ composed of quadratic forms of the metric perturbation. 

These expanded equations can then be solved order by order in the perturbation through Green function methods, where the integral is over the past lightcone of Minkowski space for $x \in M$. When working at sufficiently high PN order, the resulting integrals can be formally divergent, but these pathologies can be bypassed or cured through asymptotic matching methods (as in Will's method of the direct integration of the relaxed Einstein equations~\cite{Will:1996zj,Will:1999dq,Pati:2000vt}) or through regularization techniques (as in Blanchet and Damour's Hadamard and dimensional regularization approach~\cite{Blanchet:2000nu,Damour:2001bu}). All approaches to cure these pathologies have been shown to lead to exactly the same end result for the metric perturbation. 
  
The metric perturbation is solved for order by order, where at each order one uses the previously calculated information in the expression for $N^{\alpha \beta}$ and also to find the motion of the matter sources, thus leading to an improved expression for $T^{\alpha \beta}$ at each order. In particular, the emission of gravitational waves by a binary system causes a change in the period of that system, and this change was used by Hulse and Taylor to {\emph {indirectly}} detect gravitational waves through their observations of the binary pulsar~\cite{Hulse:1974eb}. 
In this way, the PN iterative procedure provides a perturbative approximation to the solution to the Einstein equations to a given order in the feebleness of the gravitational interaction and the speed of the bodies. 

Little work has gone into studying the mathematical properties of the resulting perturbative series. Clearly, the PN approximation should not be valid when the speed of the bodies becomes comparable to the speed of light or when the objects described are black holes or neutron stars with significant self-gravity. Damour, however, has shown that the latter can still be described by the PN approximation up to a given order in perturbation theory~\cite{Damour:1984rbx}. Moreover, recent numerical simulations of the merger of binary black holes and neutron stars have shown that the PN approximation is accurate even quite late in the inspiral, when the objects are moving at close to a third of the speed of light~\cite{Blanchet:2002xy,Boyle:2007ft,Yunes:2008tw,Boyle:2008ge,Boyle:2009dg,Zhang:2011vha}.   
  

\subsection{Resummations of the \\ PN Approximation}

The accuracy of the approximate solutions can be improved by applying resummation techniques: the rewriting of the perturbative expansion in a new form (e.g.~a Chebyshev decomposition or a Pad\'e series) that makes use of some physical feature one knows should be present in the exact solution. For example, one may know (through symmetry arguments or by taking certain limits) that some exact result contains a first-order pole at a certain spacetime position, so one could rewrite the approximate solution as a Pad\'e approximant that makes this pole explicit~\cite{Gupta:2000zb}. 
  
A particular resummation of the PN approximation that has been highly successful at approximating numerical solutions is the \emph{effective one-body approach} introduced by Buonanno and Damour \cite{bd1,bd2}. 
Recall that in Newtonian
gravity, the motion of masses $m_1$ and $m_2$ under their mutual
gravitational attraction is mathematically equivalent to the motion of a single
mass $\mu$ in the gravitational field of a stationary mass $M$, where
$M=m_1+m_2$ and $\mu=(m_1 m_2)/M$.  The effective one body
approach similarly attempts to recast the motion of two black holes
under their mutual gravitational attraction as the motion
of a single object in a given
spacetime metric.

More precisely, one recasts the two-body problem onto the problem of an effective body that moves on an effective external metric through an energy map and a canonical transformation.  The dynamics of the effective body are then described through a (conservative) improved Hamiltonian and a (dissipative) improved radiation-reaction force. The improved Hamiltonian is resummed through two sets of square-roots of PN series, in such a way so as to reproduce the standard PN Hamiltonian when Taylor expanded about weak-field and slow-velocities. The improved radiation-reaction force is constructed from quadratic first-derivatives of the gravitational waves, which in turn are product-resummed using the Hamiltonian (from knowledge of the extreme mass-ratio limit of the PN expansion) and a field-theory resummation of certain tail-effects. 

Once the two-body problem has been reformulated, the Hamilton equations associated with the improved Hamiltonian and radiation-reaction force are solved numerically, a significantly easier problem than solving the full Einstein equations. This resummation, however, is not enough because the improved Hamiltonian and radiation-reaction force are built from finite PN expansions. The very late inspiral behavior of the solution can be corrected by adding calibration coefficients (consistent with PN terms not yet calculated) to the Hamiltonian and the radiation-reaction force, which are then determined by fitting to a set of full, numerical relativity simulations (see e.g.~\cite{Buonanno:2009qa,Pan:2009wj}). 
  

The calibrated effective-one-body waveforms described above are incredibly accurate representations of the gravitational waves emitted in the inspiral of compact objects, up to the moment when the black holes merge~\cite{Babak:2016tgq}.  They become accurate after the merger by adding on information from black hole perturbation theory that we describe next~\cite{Pan:2011gk,Pan:2013rra}. 


\subsection{Perturbations about a black hole background}
\label{BH-pert-theory}

After black holes merge, they form a single distorted black hole that sheds its distortions by emitting gravitational waves and eventually settling down to a Kerr black hole.  This ``ringdown'' phase is described using perturbation theory with the Einstein vacuum equations linearized around a Kerr black hole background.  
Teukolsky showed how to obtain a wave type equation for these perturbed Weyl tensor components from the Einstein vacuum equations~\cite{saulkerr}.  The result of the Teukolsky method is that the distortions can be expanded in modes, each of which has a characteristic frequency and exponential decay time~\cite{chandra}.  The ringdown is well approximated by the most slowly decaying of these modes. This ringdown waveform can be stitched to the effective-one-body inspiral waveforms to obtain a complete description of the gravitational waves emitted in the coalescence of black holes.

\section{Mathematics and \\ Numerics}
\label{mathnum}

In numerical relativity, one creates simulations of the Einstein field equations using a computer.  This is needed when no other method will work, in particular when gravity is very strong and highly dynamical (as it is when two black holes merge).  

The Einstein field equations, like most of the equations of physics, are differential equations, and the most straightforward of the techniques for simulating differential equations are finite difference equations~\cite{numrecipes}.  In the
one dimensional setting, one approximates a function 
$f(x)$ by its values on equally spaced points 
$$
{f_i}=f(i\delta) \textrm{ for }i \in {\mathbb{N}}.
$$  
One then approximates derivatives of $f$ using differences 
$$f' \approx ({f_{i+1}}-{f_{i-1}})/(2\delta)$$
and 
$$f''(i) \approx ({f_{i+1}}+{f_{i-1}}-2{f_i})/{\delta^2}.$$
For any pde with an initial value formulation one replaces the fields by their values on a spacetime lattice, and the field equations by finite difference equations that determine the fields at time step $n+1$ from their values at time step $n$.  Thus the Einstein vacuum equations are written as
difference equations where the step 0 information is the initial data set.

One then writes a computer program that implements this determination and runs the program.  Sounds simple, right?  So what could go wrong?  Quite a lot, actually.  It is best to think of the solution of the finite difference equation as something that is supposed to converge to a solution of the differential equation in the limit as the step size $\delta$ between the lattice points goes to zero.  But it is entirely possible that the solution does not converge to anything at all in this limit.  In particular, the coordinate invariance of general relativity allows one to express the Einstein field equations in many different forms, some of which are not strongly hyperbolic.  Computer simulations of these forms of the Einstein field equations generally do not converge. 

Another problem has to do with the constraint equations.  Recall that initial data have to satisfy constraint equations.  It is a consequence of the theorem of Choquet-Bruhat that if the initial data satisfy those constraints then the results of evolving those initial data continues to satisfy the constraints.  However, in a computer simulation the initial data only satisfy the finite difference version of the constraints and therefore have a small amount of constraint violation.  The field equations say that data with zero constraint violation evolve to data with zero constraint violation.  But that still leaves open the possibility (usually realized in practice) that data with small constraint violation evolve in such a way that the constraint violation grows rapidly (perhaps even exponentially) and thus destroys the accuracy of the simulation.

Finally there is the problem that these simulations deal with black holes, which contain spacetime singularities.  A computer simulation cannot be continued past a time where a slice of constant time encounters a spacetime singularity.  Thus either the simulations must only be run for a short amount of time, or the time slices inside the black hole must somehow be ``slowed down'' so that they do not encounter the singularity.  But then if the time slice advances slowly inside the black hole and rapidly outside it, this will lead to the slice being stretched in such a way as to lead to inaccuracies in the finite difference approximation.  

Before 2005 these three difficulties were insurmountable, and none of the computer simulations of colliding black holes gave anything that could be used to compare with observations.  Then suddenly in 2005 all of these problems were solved by Frans Pretorius~\cite{pret1} who produced the first fully successful binary black hole simulation.  Then later that year the problem was solved again (using completely different methods!) by two other groups: one consisting of Campanelli, Lousto, Marronetti, and Zlochower~\cite{clz} and the other of Baker, Centrella, Choi, Koppitz, and van Meter~\cite{bakeretal}.  Though the methods are different, both sets of solutions can be thought of as consisting of the ingredients  {\it hyperbolicity, constraint damping}, and {\it excision}, and we will treat each one in turn.

{\it Hyperbolicity}. Since one needs the equations to be strongly hyperbolic, one could perform the simulations in harmonic coordinates.  However, one also needs the time coordinate to remain timelike, so instead Pretorius used generalized harmonic coordinates (as first suggested by Friedrich) where the coordinates satisfy a wave equation with a source.  The other groups implemented hyperbolicity by using the BSSN equations~\cite{bssn1,bssn2,bssn3} (named for its inventors: Baumgarte, Shapiro, Shibata, and Nakamura).  These equations decompose the spatial metric into a conformal factor and a metric of unit determinant and then evolve each of these quantities separately, adding appropriate amounts of the constraint equations to convert the spatial Ricci tensor into an elliptic operator.

{\it Constraint damping}. Because the constraints are zero in exact solutions to the theory, one has the freedom to add any multiples of the constraints to the right-hand side of the field equations without changing the class of solutions to the field equations.  In particular, with clever choices of which multiples of the constraints go on the right-hand side, one can arrange that in these new versions of the field equations small violations of the constraints get smaller under evolution rather than growing.  Carsten Gundlach and his collaborators showed how to do this for evolution using harmonic coordinates~\cite{gundl}, and their method was implemented by Pretorius.  The BSSN equations already have some rearrangement of the constraint and evolution equations.  The particular choice of lapse and shift ($\Phi$ and $X$ from eqns. (\ref{evol1}-\ref{evol2})) used by the other groups (called 1+log slicing and Gamma driver shift) were found to have good constraint damping properties. 

{\it Excision}. Because nothing can escape from a black hole, nothing that happens inside can have any influence on anything that happens outside.  Thus in performing computer simulations of colliding black holes, one is allowed to simply excise the black hole interior from the computational grid and still obtain the answer to the question of what happens outside the black holes.  By excising, one no longer has to worry about singularities or grid stretching.  Excision was first proposed by Unruh and Thornburg~\cite{thornburg}, and first implemented by Seidel and Suen and their collaborators~\cite{seidelsuen}, and used in Pretorius' simulations.  The other groups essentially achieve excision by other methods.  They use a ``moving puncture method'' which involves a second asymptotically flat end inside each black hole, that is compactified to a single point that can move around the computational grid.  The region between the puncture and the black hole event horizon undergoes enormous grid stretching, so that effectively only the exterior of the black hole is covered by the numerical grid. 

Since 2005, many simulations of binary black hole mergers have been performed, for various black hole masses and spins.   Some of the most efficient simulations are done by the SXS collaboration using spectral methods instead of finite difference methods~\cite{saulwaveform}.  (SXS stands for ``Simulating eXtreme Spacetimes'' and the collaboration is based at Cornell, Caltech, and elsewhere).  Spectral methods use the grid values $f_i$ to approximate the function $f(x)$ as an expansion in a particular basis of orthogonal functions.  The expansion coefficients and the derivatives of the basis functions are then used to compute the derivatives of $f(x)$.  Compared to finite difference methods, spectral methods can achieve a given accuracy of the derivatives with significantly fewer grid points. 


\section{Gravitational Wave \\ Experiment}
\label{experiment}

The experimental search for gravitational waves started in the 1960s through the construction of \emph{resonant bar detectors}~\cite{weber}. The latter essentially consist of a large (meter-size) cylinder in a vacuum chamber that is isolated from vibrations. When a gravitational wave at the right frequency interacts with such a bar, it can excite the latter's resonant mode, producing a change in length that one can search for. In 1968, Joseph Weber announced that he had detected gravitational waves with one such resonant bar. The sensitivity of Weber's resonant bar to gravitational waves was not high enough for this to be possible and other groups could not reproduce his experiment. 

\begin{figure}[htb]
\begin{center}
\makebox{  \includegraphics[scale=0.6]{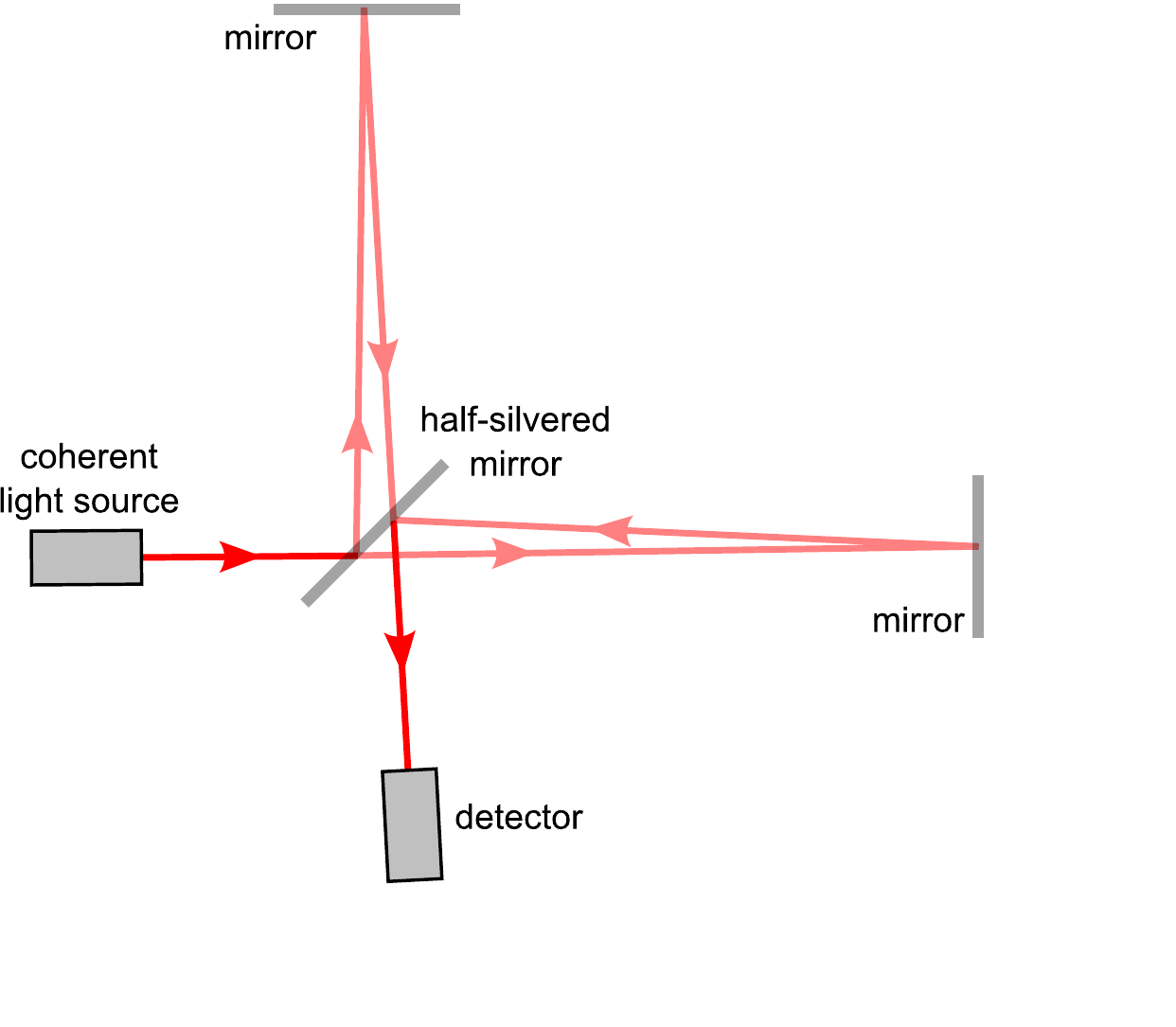}} 
\caption{Schematic diagram of a Michelson interferometer, which is at the heart of the instrumental design used by aLIGO. In the diagram, a laser beam is split into two sub-beams that travel down orthogonal arms, bounce off mirrors, and then return to recombine.}
\label{detector}
\end{center}
\end{figure}

In the late 1960s and early 1970s, the search for gravitational waves with laser interferometers began through the pioneering work of Rainer Weiss at MIT~\cite{weiss1} and Kip Thorne~\cite{pressthorne} and Ronald Drever at Caltech~\cite{drtvw}, among many others. The basic idea behind interferometry is to split a laser beam into two sub-beams that travel down orthogonal arms, bounce off mirrors, and then return to recombine. If the light travel time is the same in each sub-beam, then the light recombines constructively, but if a gravitational wave goes through the detector, then the light travel time is not the same in each arm and interference occurs. Gravitational wave interferometers are devices that use this interference process to measure small changes in light travel time very accurately so as to learn about the gravitational waves that produced them, and thus, in turn, about the properties of the source of gravitational waves. 

\begin{figure}[htb]
\begin{center}
\makebox{  \includegraphics[scale=0.5]{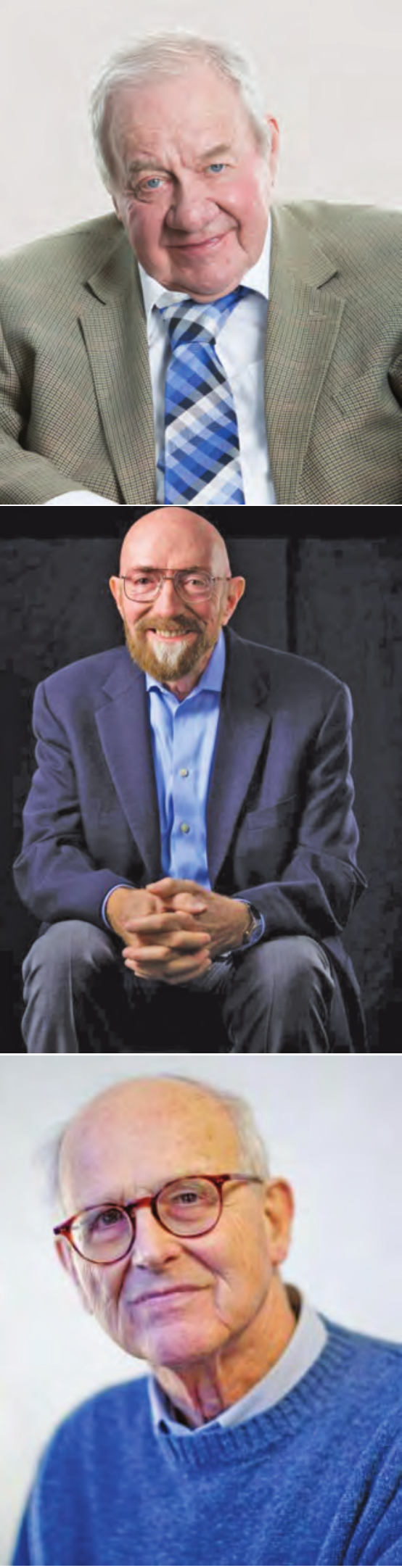}}
\caption{(from top to bottom) Ronald Drever, Kip Thorne, and Rainer Weiss pioneered the effort to detect gravitational waves with laser interferometers.}
\label{RonaldKipRai}
\end{center}
\end{figure}

The initial Laser Interferometer Gravitational-Wave Observatory (iLIGO) was funded by the National Science Foundation in the early 1990s and operations started in the early 2000s. There are actually two LIGO facilities (one in Hanford, Washington, and one in Livingston, Louisiana) in operation right now, with an Italian counterpart (Virgo) coming online soon, a Japanese counterpart (KAGRA) coming online by the end of the decade, and an Indian counterpart (LIGO-India) coming online in the 2020s. The reason for multiple detectors is to achieve redundancy and increase the confidence of a detection by observing the signal by independent detectors with uncorrelated noise. Although iLIGO was over four orders of magnitude more sensitive than Weber's original instrument in a wide frequency band, no gravitational waves were detected. 

\begin{figure}[htb]
\begin{center}
\includegraphics[width=0.5\textwidth]{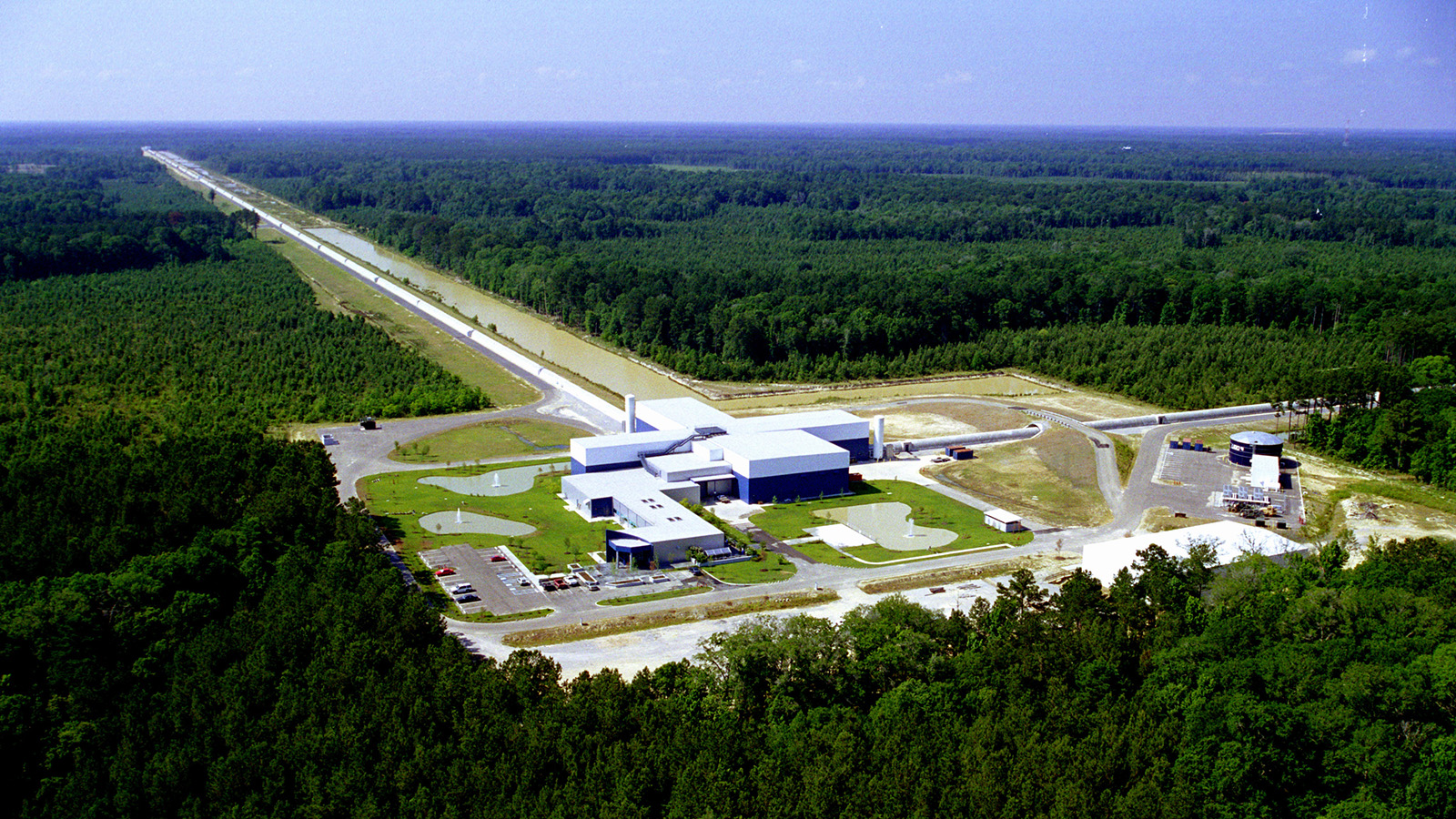}
\caption{One of the two aLIGO facilities in Livingston, Louisiana, where the interference pattern associated with a gravitational wave produced in the merger of two black holes was recorded within days of the first science run.}
\label{aLIGO}
\end{center}
\end{figure}

In the late 2000s, upgrades to convert iLIGO into advanced LIGO (aLIGO) commenced. These upgrades included an increase in the laser power to reduce quantum noise, larger and heavier mirrors to reduce thermal and radiation pressure noise, better suspension fibers for the mirrors to reduce suspension thermal noise, among many other improvements. aLIGO commenced science operations in 2015 with a sensitivity roughly $3$-$4$ times greater than that of iLIGO's last science run. 
  
Within days of the first science run, the aLIGO detectors recorded the interference pattern associated with a gravitational wave produced in the merger of two black holes 1.3 billion light years away~\cite{Abbott:2016blz}. The signal was so loud (relative to the level of the noise) that the probability that the recorded event was a gravitational wave was much larger than $5\sigma$, meaning that the probability of a false alarm was much smaller than $10^{-7}$. There is no doubt that this event, recorded on 14th September 2015, as well as a second one, detected the day after Christmas of that same year~\cite{Abbott:2016nmj}, were the first direct detections of gravitational waves. 

\begin{figure}[htb]
\begin{center}
 \includegraphics[width=\columnwidth]{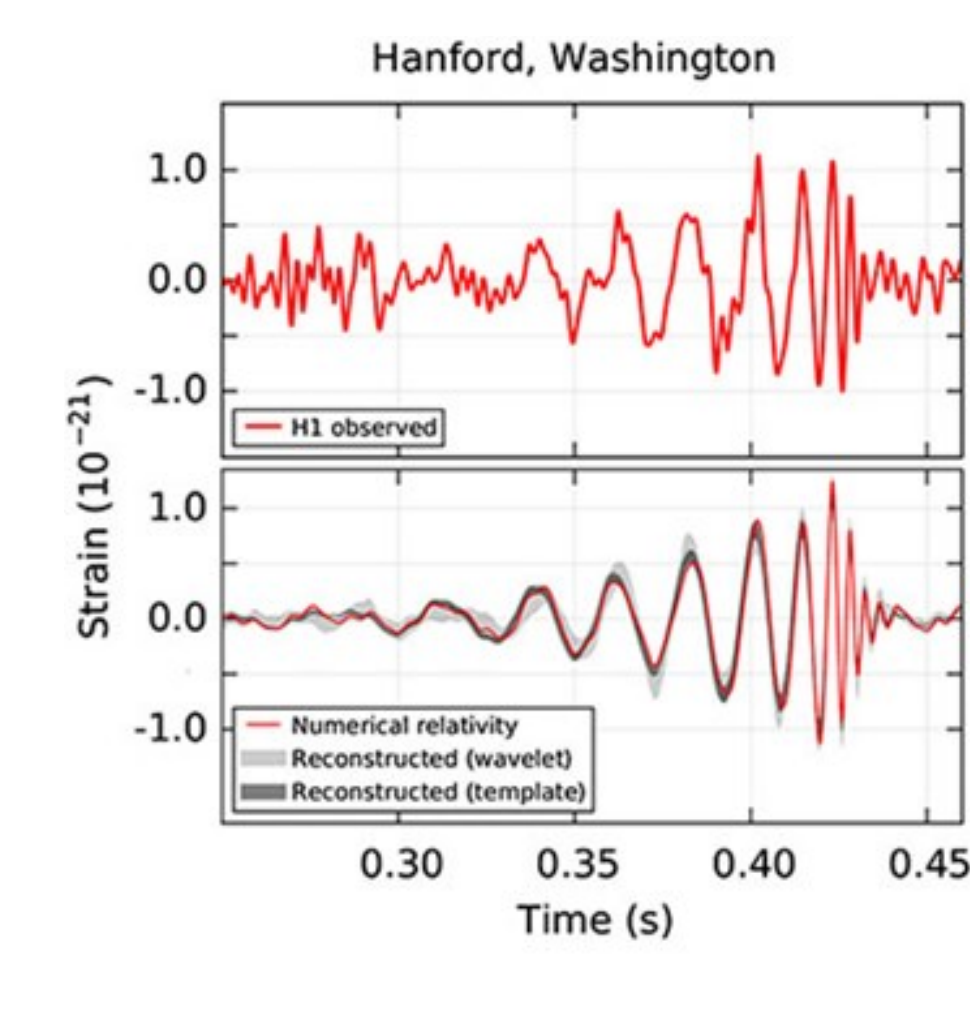}
\caption{Top: Filtered GW strain as a function of time detected at the Hanford location of aLIGO. Bottom: Best fit reconstruction of the signal using a numerical relativity simulation (red), an analytical waveform template (gray) and a set of Morley wavelets. The latter two are shown as 90\% confidence regions, while the simulation is a particular run with a choice of parameters within this the 90\% confidence region.}
\label{signal}
\end{center}
\end{figure}

In order to understand how gravitational waves are detected, we must understand how the waves affect the motion of the parts of the interferometer.  The mirrors are suspended from wires like a pendulum, but this means that for short time motion in the horizontal direction, the motion of each mirror can be treated as a spacetime geodesic. But the interferometer measures distance between the mirrors, so what we want to know is how does this distance change under the influence of a gravitational wave. The answer to this question comes from the Jacobi equation: the relative acceleration of nearby geodesics is equal to the Riemann tensor times the separation of those geodesics.  

Thus if at any time we want to know the separation, we need to integrate the Jacobi equation twice with respect to time.  However, the Riemann tensor is the second derivative of the $TT$ gauge metric perturbation.  Thus, by using this particular gauge we can say that LIGO directly measures the metric perturbation by using laser interferometry to keep track of the separation of its mirrors. \\

\section{Astrophysics and \\ Fundamental Physics}
\label{astro}

Up until now, we have created a picture of the universe from the information we have obtained from amazing telescopes, such as Chandra in the X-rays, Hubble in the optical,  Spitzer in the infrared, WMAP in the microwave, and Arecibo in the radio frequencies. This information was provided by light that traveled from  astrophysical sources to Earth. Every time humankind built a new telescope that gave us access to a new frequency range of the light spectrum, amazing discoveries were made; case in point, accretion disk signatures of black holes using X-ray astronomy. This expectation is especially true for gravitational wave detectors, which do not just open a new frequency range, but rather aim to \emph{listen to the universe} in an entirely new way: with gravity instead of light.

This new type of astrophysics has an immense potential to truly revolutionize science because gravitational waves can provide very clean information about their sources. Unlike light, gravitational waves are very weakly coupled to matter, allowing gravitational waves to go right through the intermediate matter (which would absorb light) and provide a clean picture (or soundtrack) of astrophysical sources that until now had remained obscure. Of course, this is a double-edged sword because the detection of gravitational waves is extremely challenging, requiring the ability to measure distances that are as small as $10^{-3}$ times the size of a proton over a 4 km baseline. 

The aLIGO detectors achieved just that, providing humanity with not only the first direct detection of gravitational waves, but also the first direct evidence of the existence of black hole binaries and their coalescence. As of the writing of this article, aLIGO had detected two events, both of which correspond to the coalescence of binary black hole systems in a quasi-circular orbit. Fitting the hybrid analytic and numerical models described in Sections 4 and 5 to the data, the aLIGO collaboration found that the first event consisted of two black holes with masses~\cite{Abbott:2016blz} 
 $$(m_{1},m_{2}) \approx (36.2,29.1) M_{\odot},$$ where $M_{\odot}$ is the mass of our sun, colliding at roughly half the speed of light to produce a remnant black hole with mass $$m_{f} \approx 62.3 M_{\odot}$$ and dimensionless spin angular momentum $$|\vec{S}|/m_{f}^{2} \approx 0.68,$$ located $420$ mega-parsecs away from Earth (roughly 1.3 billion times the distance light travels in one year). The second event consisted of lighter black holes, with masses~\cite{Abbott:2016nmj} $$(m_{1},m_{2}) \approx (14.2,7.5) M_{\odot}$$ that collided to produce a remnant black hole with mass $$m_{f} \approx 20.8 M_{\odot}$$ and dimensionless spin angular momentum $$|\vec{S}|/m_{f}^{2} \approx 0.74,$$ located $440$ mega-parsecs away from Earth. In both cases, the peak luminosity radiated was in the range of $10^{56}$ ergs/s with the systems effectively losing $3 M_{\odot}$ and $1 M_{\odot}$ respectively in less than $0.1$ seconds. Thus, for a very brief moment, these events produced more energy than all of the stars in the observable universe put together.

Perhaps one of the most interesting inferences one can draw from such events is that black holes (or at the very least, objects that look and ``smell'' a lot like black holes) truly do form binaries and truly do merge in nature within an amount of time smaller than the age of the universe. Until now, we had inferred the existence of black holes by either observing how other stars orbit around supermassive ones at the center of galaxies or by observing enormous disks of gas orbit around stellar mass black holes and the X-rays emitted as some of that gas falls into the black hole. The aLIGO observations are the first direct observation of radiation produced by binary black holes themselves through the wave-like excitations of the curvature they generate when they collide.   Not only did the aLIGO observation prove the existence of binary black holes, but even the first observation brought about a surprise: the existence and merger of black holes in a mass range that had never been observed before.

The aLIGO observations have demonstrated that General Relativity is not only highly accurate at describing gravitational phenomena in the solar system, in binary pulsar observations, and in cosmological observations, but also in the late inspiral, merger and ringdown of black hole binaries~\cite{Yunes:2013dva,TheLIGOScientific:2016src,Yunes:2016jcc}.  Gravity is truly described by Einstein's theory even in the most \emph{extreme gravity} scenarios: when the gravitational interaction is strong, highly non-linear, and highly dynamical.  Such consistency with Einstein's theory has important consequences on theories that modify gravity in hopes of arriving at a quantum gravitational completion. Future gravitational wave observations will allow us to verify many other pillars of Einstein's theory, such as that the gravitational interaction is parity invariant, that gravitational waves propagate at the speed of light, and that it only possesses two transverse polarizations.

The detection of gravitational waves is not only a spectacular confirmation of Einstein's theory, but also the beginning of a new era in astrophysics. Gravitational waves will provide the \emph{soundtrack } to the movie of our universe, a soundtrack we had so far been missing with telescopes. No doubt that they will be a rich source for new questions and inspiration in physics as well as mathematics. We wait anxiously for the unexpected beauty this music will provide. 

\section*{Acknowledgments}

The authors acknowledge their NSF support. 
LB is supported by NSF CAREER grant DMS-1253149 to The University of Michigan. DG is supported by NSF grant PHY-1505565 to Oakland University. N.Y. acknowledges support from NSF CAREER Grant PHY-1250636. 
   \\ \\

\section*{Nobel Prize in Physics 2017}

While we were editing the arxiv version of this article, \footnote{See the official announcement on the Nobel Prize website 
\url{https://www.nobelprize.org/nobel_prizes/physics/laureates/2017/press.html}} ``the Royal Swedish Academy of Sciences has decided to award the Nobel Prize in Physics 2017 with one half to Rainer Weiss, LIGO/VIRGO Collaboration, 
and the other half jointly to Barry C. Barish, LIGO/VIRGO Collaboration, and Kip S. Thorne, LIGO/VIRGO Collaboration, 
for decisive contributions to the LIGO detector and the observation of gravitational waves."

\begin{figure}[htb]
\begin{center}
\makebox{  
\includegraphics[scale=0.5]{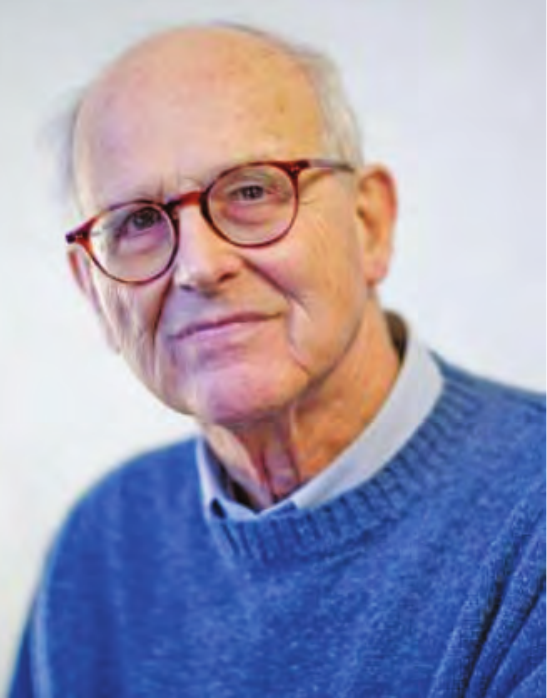} 
}
\caption{ Rainer Weiss (LIGO/VIRGO Collaboration).}
\label{Rai}
\end{center}
\end{figure}

\vspace{-1cm}

\begin{figure}[htb]
\begin{center}
\makebox{  
\includegraphics[scale=0.15]{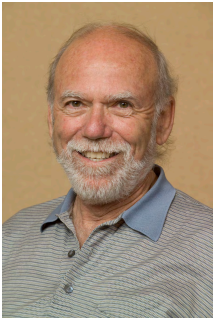}
}
\caption{Barry Barish (LIGO/VIRGO Collaboration).}
\label{Barry}
\end{center}
\end{figure}

\vspace{-1cm}

\begin{figure}[htb]
\begin{center}
\makebox{  
\includegraphics[scale=0.5]{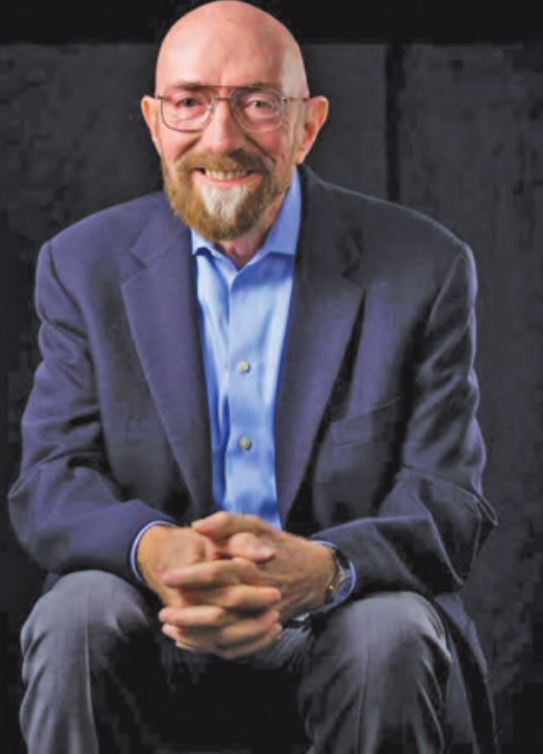} 
}
\caption{Kip Thorne (LIGO/VIRGO Collaboration).}
\label{Kip}
\end{center}
\end{figure}

 
%
\vspace{5pt}

{\scshape \small Photo Credits} 

{\small Figures \ref{Einstein} and \ref{Einstein2} are in the public domain.} 

{\small Figure \ref{detector} is courtesy of Ed Stanner. } 

{\small Figures \ref{RonaldKipRai}, \ref{Rai}, and \ref{Kip} are courtesy of the Gruber Foundation. } 

{\small Figure \ref{aLIGO} is courtesy of the Caltech/MIT/LIGO Lab.} 

{\small Figure \ref{signal} is courtesy of \cite{ligo1} B.P. Abbot et al. ``Observation of Gravitational Waves from Binary Black Hole Merger". PRL. 116. 061102. (2016). } 

{\small Figure \ref{Barry} is in the public domain. } \\

\vspace{5cm}

{\scshape Lydia Bieri \\
Department of Mathematics \\ 
University of Michigan \\ 
Ann Arbor, MI 48109, USA } \\ 
Email address: lbieri@umich.edu \\ \\ 

{\scshape David Garfinkle \\ 
Physics Department \\
Oakland University \\ 
Rochester, MI 48309, USA, \\ and \\ 
Michigan Center for Theoret. Physics \\ 
Randall Laboratory of Physics \\ 
University of Michigan \\ 
Ann Arbor, MI 48109, USA} \\ 
Email address: garfinkl@oakland.edu \\ \\ 

{\scshape Nicol\'as Yunes \\ 
eXtreme Gravity Institute \\
 Department of Physics \\ 
 Montana State University \\
 Bozeman, MT 59717 USA } \\ 
Email address: nyunes@physics.montana.edu
\end{document}